\documentclass[prl,superscriptaddress,twocolumn,showpacs,preprintnumbers,psfig,aps,epsf,amsmath,amssymb]{revtex4}
\usepackage{graphicx}
\usepackage{dcolumn}
\usepackage{bm}
\newcommand{\boldtau}{\mbox{\boldmath $\tau$}}

\newcommand{\boldalpha}{\mbox{\boldmath$\alpha$}}
\newcommand{\bolddelta}{\mbox{\boldmath$\delta$}}

\newcommand{\boldnabla}{\mbox{\boldmath$\nabla$}}
\newcommand{\boldxi}{\mbox{\boldmath$\xi$}}

\newcommand{\We}{\textit{We}}
\begin{document}

\title{Toward a structural understanding of turbulent drag reduction:
nonlinear coherent states in viscoelastic shear flows}

\author{Philip A. Stone}
\affiliation{%
Department of Chemical Engineering
}%
\author{Fabian Waleffe}
\affiliation{%
Departments of Mathematics and Engineering Physics, University of Wisconsin-Madison,
Madison, WI 53706. 
}%
\author{Michael D. Graham}
\email{graham@engr.wisc.edu}
\affiliation{%
Department of Chemical Engineering
}%
\date{\today}

\begin{abstract}

Nontrivial steady flows have recently been found that capture
the main structures of the turbulent buffer layer.  We study the
effects of polymer addition on these ``exact coherent states" (ECS) in
plane Couette flow.  Despite the simplicity of the ECS flows,
these effects closely mirror those observed experimentally:
structures shift to larger length scales, wall-normal fluctuations are
suppressed while streamwise ones are enhanced, and drag is reduced. 
The mechanism underlying these effects is elucidated.  These results 
suggest that the ECS are closely related to 
buffer layer turbulence.
\end{abstract}


\maketitle
\thispagestyle{empty}

Rheological drag reduction, the suppression by additives of skin
friction in turbulent flow, has received much attention since its
discovery in 1947~\cite{virk75,lumley69,mccomb}.  For many
polymer-solvent systems, the pressure drop measured in the pipe flow
of the solution can be $30-50\%$ less than for the solvent alone. 
The
central rheological feature of drag-reducing additives is their
extensional behavior in solution: for dilute polymer solutions in
particular the stresses arising in extensional flow can be orders of
magnitude larger than those developed in a shear flow.  This fact is
well-recognized; nevertheless the mechanism of interaction between
polymer stretching and turbulent structure is not well-understood and
the goal of the present work is to attempt to shed light on this
interaction.

A key structural observation from experiments and direct numerical 
simulations (DNS) of
drag-reducing solutions is the modification of the buffer region near
the
wall~\cite{tiederman90,tiederman72,nieuwstadt97,nieuwstadt98,smith99,beris97,beris01}.
 It has long been known that the flow in this region is very
structured, containing streamwise vortices that lead to streaks in the
streamwise velocity~\cite{robinson91}; these structures are thickened in
both the wall-normal and spanwise directions during flow of drag
reducing solutions~\cite{tiederman90,tiederman72}.  Because of its
importance in the production and dissipation of turbulent
energy~\cite{robinson91}, any effort to mechanistically understand
rheological drag reduction should address this region.

 To better understand the effect of the polymer on the buffer layer,
 we wish to study a model flow that has structures similar to those
 seen in this region but without the full complexities of
 time-dependent turbulent flows.  Fortunately, a family of such 
 flows exists, in the recently-discovered ``exact coherent states''
 (ECS) found by computational bifurcation analysis in plane Couette
 and plane Poiseuille
 flows~\cite{nagata90,busse97,waleffe98,eckhardt98,waleffe01}.  These
 are three-dimensional, traveling wave flows (hence steady in a
 traveling reference frame)  that appear via saddle-node
 bifurcations 
 \footnote{A saddle-node bifurcation, also known as a 
 turning point, limit point or fold, arises 
when two steady states appear spontaneously as a parameter increases \cite{Strogatzbook}.
} 
at a Reynolds number somewhat below the transition value seen in
 experiments~\cite{daviaud92,dauchot98}.  The structure of the ECS
 captures the counter-rotating staggered streamwise vortices that
 dominate the structure in the buffer region.  From the dynamical
 point of view, there is evidence that these states form a part of the
 dynamical skeleton of the turbulent flow: i.e., they are saddle
 points that underlie the strange attractor of
 turbulence~\cite{jimenez01,kida01}.  Finally, the nonlinear
 self-sustaining mechanism underlying these states has been
 elucidated~\cite{waleffe97}.  A perturbation of the base flow in the
 form of streamwise vortices redistributes the streamwise momentum of
 the flow.  This redistribution creates spanwise fluctuations in the
 streamwise velocity, the ``streaks''.  The spanwise inflections in
 the streamwise velocity profile lead to a three-dimensional
 instability that develops into staggered nearly-streamwise vortices
 that regenerate the streaks.  Because the ECS capture the structures
 of the buffer region and are mechanistically well-understood, we
 believe that they provide an excellent simplified, yet still exact,
 model flow for studying polymer drag reduction.  The leading order
 effect of viscoelasticity on the ECS is therefore the focus of the
 present study.

To begin, we briefly 
describe a general result relating polymer stretch to flow kinematics. 
For a trajectory in a flow field, the Liapunov exponents give the
Lagrangian time-averaged rate of stretch of material lines.  If the
largest Liapunov exponent, ${\sigma}_{max}$, is positive, the flow is
extensional on average.  In particular, for homogeneous turbulence, the expected
value of ${\sigma}_{max}$ is positive~\cite{pope90} and we show below
that this is also the case for the ECS. Now consider the dynamics of a
Hookean dumbbell model of a polymer in a flow field.  The end-to-end
vector ${\bf q}$  of the dumbbell evolves in the flow field, ${\bf v}$, as:
\begin{equation}
\frac{D{\bf q}}{Dt} = {\bf q}{\cdot}{\boldnabla}{\bf v} - \frac{1}{2 \lambda}{\bf q} + \boldxi(t),
\end{equation}
where $D/Dt$ is the time derivative evaluated on a fluid element,
$\lambda$ is the stress relaxation time for the dumbbell and
$\boldxi(t)$ is the random Brownian force.  Noting that an
infinitesimal material line satisfies the same expression but with
$\lambda^{-1}$ and $\boldxi$ set to zero, it is straightforward to
show that Hookean dumbbells will stretch indefinitely in a flow if and
only if
$
 {\lambda}{\sigma}_{max} \equiv \We_{\sigma}  > \frac{1}{2},
$
~where $\We_{\sigma}$ is a Weissenberg number based on ${\sigma}_{max}$. 
This result is a specific statement of an idea that originated with
Lumley~\cite{lumley72} (see also \cite{Ryskin87,leal90,chertkov00,lebedev00}).  The  
computations below show the importance of $\We_{\sigma}$ in 
determining the effect of polymers on coherent structures.

We study here the effect of polymer on the exact coherent states that
arise in a variant of plane Couette flow \cite{waleffe98}.  Denoting the
streamwise direction as $x$, the wall-normal direction as $y$, and the
spanwise, or vorticity, direction as $z$, we consider a flow with
boundary conditions
$
\frac{\partial v_{x}}{\partial y} = 1, 
v_{y} =  \frac{\partial v_{z}}{\partial y} = 0 {\mbox{ at }} y = \pm 1.
$~
The
characteristic velocity, $U$, and the half-height of the channel, $l$,
have been used to scale the velocity and positions, respectively. 
These ``constant vorticity'' boundary conditions provide an advantage
over no-slip conditions in that they allow us to model only the buffer
region in our domain by eliminating the viscous sublayer.  (Exact
coherent states found using no-slip BCs~\cite{waleffe01} show a
qualitatively identical vortical structure, only offset from the wall
by a small region comprising the viscous sublayer.)  Periodic boundary
conditions are applied in the streamwise and spanwise directions.  For
this study, the wavelength of the structures in the streamwise and
spanwise directions are fixed at $\ell_x = 2 \pi / 0.40$ and $\ell_z =
2 \pi / 1.0$, respectively (or 165 and 66, if expressed in wall units 
at a Reynolds number of 110).   For this flow, a trivial (Couette) base state 
exists, $v_{x}(y) =
y$; the maximum mean velocity for the ECS is significantly reduced
compared to the base state velocity due to the enhanced transport of momentum~\cite{waleffe98}.

In our formulation, time, $t$, is scaled with $l/U$, and pressure,
$p$, with ${\rho}U^2$, where ${\rho}$ is the fluid density.  The
stress due to the polymer, $\boldtau_p$, is nondimensionalized with
the polymer elastic modulus, $G = \eta_p/{\lambda}$, where $\eta_p$ is
the polymer contribution to the zero-shear viscosity and $\lambda$ is
the relaxation time for the polymer.  The momentum and mass 
balances are 
\begin{gather}
\frac{D{\bf v}}{Dt} 
=  - \boldnabla p  
+ \beta \frac{1}{Re} {\nabla^2}{\bf v} + (1-\beta)
\frac{1}{Re^2} \frac{1}{El} (\boldnabla \cdot \boldtau_p), 
\label{eq:NSE1}\\ 
\boldnabla \cdot {\bf v} = 0, \label{eq:NSE2}
\end{gather} 
where ${\eta}_s$ is the solvent viscosity, 
$El = {\lambda}(\eta_s + \eta_p)/\rho l^2$ and 
$\beta = \eta_s / (\eta_s + \eta_p)$.  The Reynolds
number, $Re$, is based on the total viscosity, $Re = {\rho}{U}{l}/{(\eta_s +
\eta_p)}$.  

We calculate the polymer stress with the commonly used FENE-P
model~\cite{dpl2}, which idealizes the polymer molecules as
bead-spring dumbbells with finitely extensible springs.  With this
model, the non-dimensional structure tensor $\boldalpha$ ($=\langle {\bf 
q}{\bf q}\rangle$, where $\langle\rangle$ denotes ensemble average) evolves 
according to: 
\begin{gather} \label{alphaequ}
\frac{\boldalpha}{1-\frac{tr \boldalpha}{b}} + \We \left( 
\frac{D {\boldalpha}}{D t}- {\boldalpha} \cdot \boldnabla {\bf v} - 
\boldnabla {\bf v}^{T}\cdot {\boldalpha} \right) = 
\frac{b\bolddelta}{b+2}, \\ \label{taupequ}
\boldtau_p = \frac{b+5}{b}\left(\frac{\boldalpha} {1-\frac{tr
\boldalpha}{b}} - \left( 1 - \frac{2}{b+2} \right) \bolddelta \right),
\end{gather}
where $\We =
\frac{{\lambda}U}{l}$ is the Weissenberg number based on the wall 
shear rate and $b$ is
proportional to the maximum extension of the dumbbell:  $tr
{\boldalpha}$ cannot exceed $b$.
A simple measure of the importance of extensional polymer stress is
the magnitude of the parameter $Ex = \frac{2}{3} \frac{b
\eta_p}{\eta_s}.  $ In uniaxial extension with extension rate
$\dot{\varepsilon}$, $Ex = 1 $ implies that ${\boldtau}_p =
{\boldtau}_v \ \mbox{as}\ \dot{\varepsilon} \rightarrow \infty $ where
${\boldtau}_v$ is the solvent contribution to the stress.  The polymer
can significantly affect the flow only when $Ex \gtrsim 1$.  In the
present flow this parameter also represents the maximum ratio of
polymer stress to Reynolds shear stress (the flux of $x$-momentum due to 
fluctuations in $y$-velocity); the scaling theory of Tabor and de
Gennes \cite{degennes86,degennes90,white00} treats the regime
$Ex \gg 1$.

The governing equations are solved through a
Picard iteration.  A given velocity field is first used to calculate
the polymer stress tensor, ${\boldtau}_p$, by time-integrating
Eq.~\ref{alphaequ} until a steady state is attained.  For the new
${\boldtau}_p$, a steady state of the momentum and continuity
equations is found by Newton iteration.  The resulting velocity field,
is used to compute the new ${\boldtau}_p$, and the process
is repeated until the velocity field converges.  
Equations~\ref{eq:NSE1}-\ref{eq:NSE2} are discretized as in \cite{waleffe98}, using a
Fourier-Galerkin formulation with typically a $7{\times}19{\times}7$
grid.  Equation~\ref{alphaequ} is discretized with a
Fourier-pseudospectral method, typically with a
$32{\times}32{\times}32$ grid, and time-integration performed with an
Adams-Bashforth method.  To achieve numerical stability, a small
diffusive term is added to Eq.~\ref{alphaequ}~(cf.~\cite{beris97}) and
integrated with a Crank-Nicholson scheme.  

\begin{figure}[htbp]
\centerline{\includegraphics{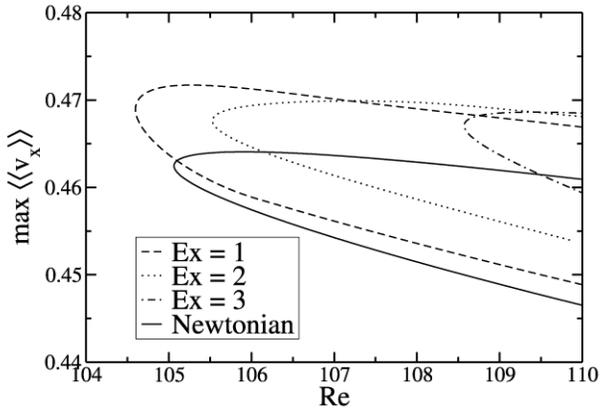}}
\caption{Newtonian and viscoelastic ($El =0.20$,
$\beta = 0.97$) bifurcation diagrams. 
Each curve 
 represents a locus of steady state ECS flows; the leftmost point on 
 each curve is the position of the saddle-node bifurcation.}
\label{bifex}
\end{figure}

Before presenting the effects of the polymer on the ECS, we recall the
result that $We_{\sigma} > 1/2$ implies large stretch of polymer
chains.  
For the Newtonian ECS  at $Re=110$ on
the lower branch of the bifurcation diagram (see Fig.~\ref{bifex}), the 
velocity field is very nearly ergodic, with ${\sigma}_{max} \approx
0.030$.  
The condition $\We_{\sigma} > 1/2$ thus translates into $\We
\gtrsim 17$ for large polymer stretch, and for $Ex = O(1)$, will
define the onset condition for the polymer to begin to strongly affect
the flow field.  In DNS of a FENE-P fluid in plane channel flow,
Sureshkumar, et al.~\cite{beris97}, found no drag reduction at $\We =
12.5$ and significant drag reduction at $\We = 25$; another recent DNS
study \cite{baron02} places the onset
value at $\We \approx 20$.  This close correspondence between the
onset condition predicted from the ECS kinematics and that found by
DNS strongly suggests that the ECS model captures the essential
structure of the buffer layer.

Fig.~\ref{bifex} shows how the addition of polymer stress affects the
bifurcation diagram (i.~e.~the locus of steady state flows) for $El =
0.20$ ($We_{\sigma} \approx 2/3$) and $Ex$ = 1 to 3.  On the 
$y$-axis of the diagram is the maximum value of the streamwise- and 
spanwise-averaged streamwise velocity $\langle\langle v_{x}\rangle\rangle$.  (The 
trivial Couette solution is at $\max\langle\langle v_{x}\rangle\rangle=1$.) When $Re$ attains a certain
value that we denote $Re_{sn}$, two new steady solutions appear in a
saddle-node bifurcation.
For small values of $Ex$, $Re_{sn}$ decreases compared to the Newtonian
value, but once the polymer stress begins to exceed the viscous ($Ex
\approx 1.5$), $Re_{sn}$ increases above the Newtonian value -- the
presence of the polymer is suppressing the ECS. Qualitatively
identical behavior is observed experimentally in the onset
Reynolds number for turbulent pipe flow \cite{nieuwstadt98}.
Note that, for a given velocity $U$ and total viscosity $\eta$, the
increase in critical Reynolds number corresponds to an increase in the
characteristic length scale of the coherent structure, again consistent
with experimental 
observations\cite{tiederman90,tiederman72,beris97}.  Finally, we see 
that $\langle\langle v_{x}\rangle\rangle$ becomes larger for the viscoelastic flows 
than for the Newtonian -- drag reduction occurs.

\begin{figure}[htbp]
\centerline{\includegraphics[width=3.25in]{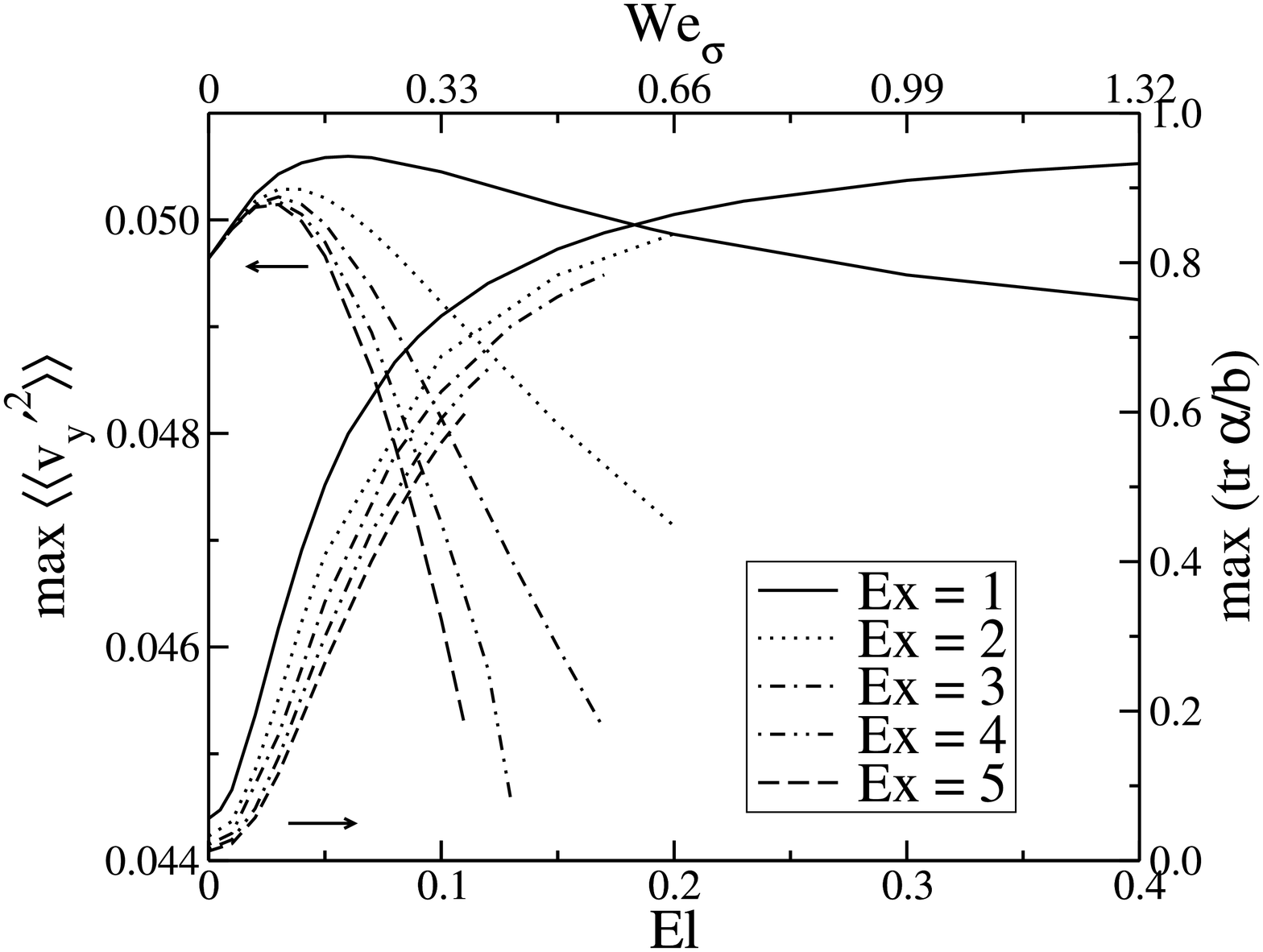}}
\caption{Wall normal velocity and maximum polymer stretch \emph{vs.}
$El$ and $\We_{\sigma}$, lower branch solutions, $Re = 110$, $\beta = 0.97$. }
\label{threeD}
\end{figure}

To examine more closely the effect of the polymer stress on the
velocity fields, Fig.~\ref{threeD} shows results at constant $Re$
while varying $El$, or, equivalently, $\We_{\sigma}$ (based on ${\sigma}_{max} = 0.030$).  Here
we plot $\max \langle\langle v_{y}'^{2} \rangle\rangle$, where the prime denotes the
fluctuating part of a quantity, here wall-normal velocity.  At
$Ex=1$, after an initial increase, $\max \langle\langle v_{y}'^{2} \rangle\rangle$
decreases below the Newtonian value and eventually saturates, as the
polymer stress asymptotes at high $\We_{\sigma}$ to a fixed value
relative to the viscous stress.  In this case, the polymer stretch
becomes nearly uniformly large throughout the domain.  The decrease in
wall-normal velocity with $\We_{\sigma}$ is even more drastic as the
extensibility parameter $Ex$ increases.  Similar trends are 
seen in the streamwise enstrophy and Reynolds shear stress.
Since $Ex$ is related to the extensional viscosity of the viscoelastic
solution, these results show the importance of extensional stresses in
affecting the ECS. The spatial maximum of $tr {\boldalpha}$, which is
proportional to the square of the polymer extension, is also presented
in Fig.~\ref{threeD}.  Finally, note that the majority of both the
polymer stretch and the change in wall-normal velocity occurs in the
range $0.1 < \We_{\sigma} < 1.0$.  In contrast to the decreases in
wall normal fluctuations, streamwise enstrophy and Reynolds shear
stress for $\We_{\sigma}\gtrsim 0.1$, the streamwise fluctuations
$\langle\langle v_{x}'^{2}\rangle\rangle$ are found to increase.  All of these
trends are observed in DNS and experiments \cite{virk75,mccomb,beris97}.

\begin{figure}[htbp]
\centerline{\includegraphics[width=3.25in]{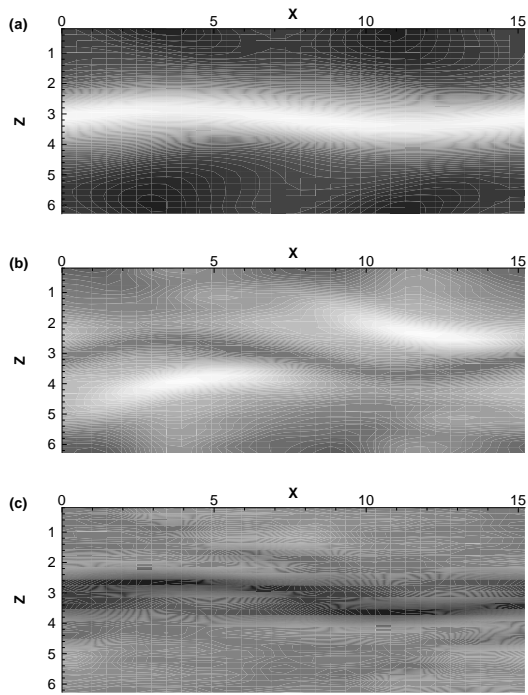}}
\caption{Streamwise velocity for the
Newtonian (N) and viscoelastic (VE) solutions and polymer force at $y
= 0.875$. Lower
branch solutions at $Re = 110$ and for the
viscoelastic solution $El= 0.15$, $Ex = 3$, and ${\beta} = 0.97$. (a)
$v_{x,N}$ (range: 0.0 (black) --- 0.889 (white)) (b) $v_{x,VE} -
v_{x,N}$ (range: -0.0395 (black) --- 0.0395 (white)) (c) $f_x$
(range: -0.00413 (black) --- 0.00413 (white)).  By symmetry, identical
behavior with opposite signed velocity is occurring at $y=-0.875$.}
\label{ycomp3D}
\end{figure}

Turning now to how, and why, the flow structure is changed by the
polymer dynamics, figures~\ref{ycomp3D}a-b show (a) the streamwise
velocity $v_{x}$ at $y=0.875$  (where the maximum in the polymer stress 
occurs) for the Newtonian lower branch solution
at $Re=110$ and (b) the difference $v_{x,VE}-v_{x,N}$ between the
viscoelastic (VE) and Newtonian (N) solutions.  Here we see
the ``streak'' (white ribbon) and -- by adding the pattern of
fig.~\ref{ycomp3D}b to that of \ref{ycomp3D}a -- that this streak is
``straightened out'' by the viscoelasticity.
Figure~\ref{ycomp3D}c shows a contour plot of $f_{x}$, the
$x$-component of the force $\boldsymbol{f}$ exerted by the polymer on
the fluid,
($\boldsymbol{f}=(1-\beta)Re^{-2}El^{-1}\boldsymbol{\nabla}\cdot{\boldtau}_p$),
at $y = 0.875$.  This force is significantly negative and corresponds
spatially to the region where fluid elements are leaving the streak to
move into the vorticity-dominated regions.  The $y$ and $z$ components
of the force have also been examined; they are smaller than the
$x$-component but are clearly seen to work directly against the vortex
motions: e.g. where $v_{y}$ is highly positive in an upwelling,
$f_{y}$ is highly negative.  This behavior is also seen in the buffer
layer structures in the DNS study of a drag-reducing polymer solution by De Angelis
\emph{et al.}~\cite{piva02}.  The origin for this structure of the
polymer force field becomes apparent on examination of polymer
stresses along fluid trajectories: polymer molecules stretch in (or
moving into) the streak regions, remaining highly stretched until they
begin to leave the streak.  As molecules move from the streak into and
around the vortices, they relax.  The spatial gradients in stress
accompanying this relaxation work against the vortices, ``unwinding''
them.  This vortex suppression leads to collapse of the self-sustained
process -- or more precisely to a shift of the process to larger
scales -- and thus to drag reduction.

To summarize, we list several points of agreement between our results
and observations from DNS and experiments in fully turbulent flow, namely: (1) the ECS bear a
strong similarity to the structures observed or educed from structural
studies of the buffer layer 
and apparently underlie
its dynamics
; (2) for $O(1)$ values of $Ex$,
the onset Weissenberg number for drag reduction predicted from the ECS
kinematics agrees closely with DNS results
; (3) the effects of viscoelasticity on the Couette flow ECS are very
similar to those observed in the buffer layer: (a) wall normal
velocity fluctuations are suppressed and streamwise ones enhanced, (b)
Reynolds shear stress decreases, (c) streamwise vorticity decreases,
(d) the velocity fluctuations and polymer force are anticorrelated,
and (e) drag is reduced.  Finally, at fixed $U$ and $\eta$ the upward
shift in the onset Reynolds number corresponds to an increase in
length scale for the structures, again consistent with experiments. 
These successes show that studying the ECS holds promise for capturing
the essential physics of drag reduction.  Indirectly, they also
validate the view that the ECS underlie Newtonian turbulence, because
the effects of polymers on the ECS so closely mirror their effects on
full turbulence.

The authors gratefully acknowledge support from NSF and the donors of
the Petroleum Research Fund, administered by the American Chemical
Society.

\newcommand{\noopsort}[1]{} \newcommand{\clarify}[1]{#1}

\end{document}